# Experimental Investigation of the Effect of Pilot Tone Modulation on Partial Response Modulation Formats

Peter Madsen[(1)], Anders T. Clausen[(1)], Annika Dochhan[(2)], Michael Eiselt[(2)]

[(1)] Technical University of Denmark Ørsteds, Plads, bygning 343,2800 Kgs.Lyngby, Denmark,
petma@fotonik.dtu.dk
[(2)] ADVA Optical Netwoking SE, 98617 Meiningen-Dreissigacker, Germany

**Abstract** *This paper presents an experimental investigation of 8% pilot tone modulation depth is a system transmitting NRZ, PAM4 and Duobinary. The penalty from the pilot tone increases with signal amplitude levels and reaches a received power penalty of 3 dB.*

## Introduction

The great number of devices that are being connected through the so-called Internet of Things (IoT) along with societies trend of using capacity hungry streaming services and lots of social media applications. All lead to ever-increasing internet data traffic. Many new transmission technologies are evolving to keep up with the heavy traffic demand, while maintaining a low-cost point. One such transmission technology is Wavelength Division Multiplexing (WDM) for Passive Optical Networks (PONs). ITU-T Study Group (SG) 15 has agreed on recommendation G.698.2[1] (ex. G.metro), describing a low cost WDM system that uses wavelength tunable Tail End Equipment (TEE) with the capability to automatically adapt its frequency to fit the channel passband of the optical multiplexers/demultiplexers. Originally this draft recommendation targets WDM application up to 10 Gbit/s in a 50 GHz channel grid spacing, also referred to as Dense WDM (DWDM). In preparation for handling future, even higher traffic, applications with data speeds up to 25 Gbit/s are now considered. When data speeds are increased, low complexity multilevel modulation formats are considered to keep the required transmission Bandwidth (BW) low. The low BW allow for the use of lower BW equipment thus making the system lower cost. One key challenge with this consideration is the use of Pilot Tone (PT) modulation[2].

In order to control the wavelength tuning of the TEE in the system, a pilot tone (PT) communication, based on envelope modulation of the optical data signal, is exchanged between the TEE and the Head End Equipment (HEE). The effect of the PT on the system performance has been investigated for 10 Gbit/s Non-Return-to-Zero (NRZ) based transmission[3]. It is expected that the PT have a larger impact on multi-level signals.

In this paper an experimental investigation of the effect of PT modulation on various multilevel modulation formats is presented. The data speed is 25 Gbit/s, and the modulation formats under test are NRZ, Duobinary, 4 level Pulse Amplitude Modulation (PAM4). The target transmission distance is up to 20 km over Standard Single Mode Fiber (SSMF) in a frequency range of 191.5-193.4 THz. During operation, the PT modulation depth ($m$) from the TEE to the HEE should be in the range of 5-8% reference. The target in this paper is 8% to evaluate the worst case impact from the PT.

## Modulation Formats

The modulation formats used in this set-up are generated using Digital Signal Processing (DSP) and transmitted as an electrical signal using an Arbitrary Waveform Generator (AWG).

Duobinary modulation is a partial response format[2] based on the principals of introducing controlled Intersymbol Interference (ISI) to a NRZ bit sequence. This can be done either by Bessel lowpass filtering of a NRZ signal or by delay and add filtering. In this paper a lowpass Bessel filter with a 3dB cutoff at ~11.5 GHz is used. The Frequency response of the filter is seen in Fig. 1. In the receiver DSP, the signal is demodulated by applying the modulo 2 operation on the received signal. Because of the introduced ISI, the required BW for duobinary modulation is lowered to almost half of that of the original NRZ signal.

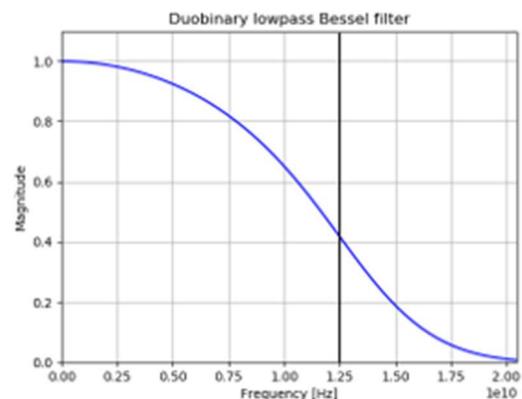

**Fig. 1:** Frequency response of the 12.5 GHz Lowpass Bessel filter.

PAM4 is normally generated by combining two consecutive bits of a bit stream into one of four amplitude levels. In this paper, the PAM4 is generated by encoding two bits of the NRZ as in

**Tab. 1:** PAM4 Gray-coding

| Bit 1 | Bit 2 | Symbol |
|-------|-------|--------|
| 0 | 0 | 0 |
| 1 | 0 | 1 |
| 1 | 1 | 2 |
| 0 | 1 | 3 |

Tab.1

The encoding results in an effectiveness of 2 bits per symbol. This corresponds to a 25 Gbit/s NRZ signal being converted to a 12.5 Gbaud PAM4 signal, effectively halving the signal bandwidth.

The PT impacts the system performance by acting as extra noise on top of the signal. When the system is in operation, the PT frequency should be between 47.5-52.5 kHz and the modulation depth should be between 5-8 %[1].

$$m = \frac{P_{max} - P_{min}}{P_{max} + P_{min}} * 100\% \quad (5)$$

Eq. (5) defines the modulation depth $m$ at the PT frequency as the peak-to-peak (p2p) variation of the average optical power, divided by twice the average optical power. The maximum PT modulation depth is measured at the optical back-2-back, after lowpass filtering at 280 kHz[1].

**Experimental Setup**

The experimental investigation of the PT effect

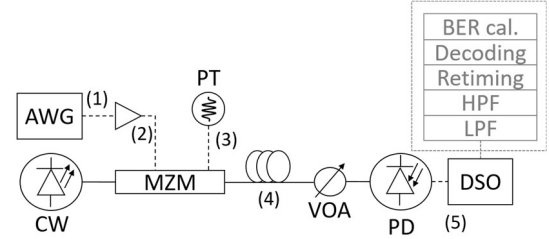

**Fig. 2:** The experimental setup. (1) The electrical output from the AWG. (2) The electrical amplification stage. (3) The PT. (4) The optical fiber length. (5) The electrical interface of the DSO

on multilevel modulation formats is realized through Bit Error Rate (BER) curves from the setup in Fig. 2.

By digital processing, a 25 Gbit/s NRZ is converted into Duobinary and PAM4 following the process from the above section. The sequences are sent to the 50-GSa/s AWG. The p2p voltage of the electrical AWG output (1) in Fig. 2 is adjusted to 500mV for NRZ and Duobinary, and to 250mV for PAM4. The reduced p2p output of the AWG, with 26 dB amplification at (2) in Fig. 2. ensures the Mach Zehnder Modulator (MZM) operates in the linear regime for PAM4. The MZM is biased by an offset PT signal. The PT is generated from a low frequency AWG delivering a sine wave at (3) in Fig. 2. The settings of the PT generator include amplitude of the sin wave and DC offset. The DC offset determines the MZM bias point and the combination of DC offset, amplitude and modulation format determines the PT modulation depth. PT modulation is adjusted to a depth of ~8 % according to reference. The required lowpass filter of 280 kHz is implemented in DSP. Continuous Wave (CW) laser with an output power of 16 dBm is used as optical source for the MZM. A Polarization maintaining fiber is used between CW and MZM. After the MZM, the signal is transmitted through a Standard Single Mode Fiber (SSMF) with a length of either 0 km (B2B) or 20 km, which is the target of G.698.2. After the fiber, a Variable Optical Attenuator (VOA) is used to enable BER measurements vs. received power. The receiver consists of a linear photo diode (PD), which is directly connected to an 80 GSa/s Digital Storage Oscilloscope (DSO) with a BW of 1-28 GHz. The DSO allows for signal processing using DSP. In DSP the signal is further lowpass filtered to remove noise from the signal. A highpass filter with cutoff at 280 kHz is implemented to remove the PT from the signal. After filtering, the signal is resynchronized and decoded depending on modulation format. The BER measurement is done using bit-by-bit comparison.

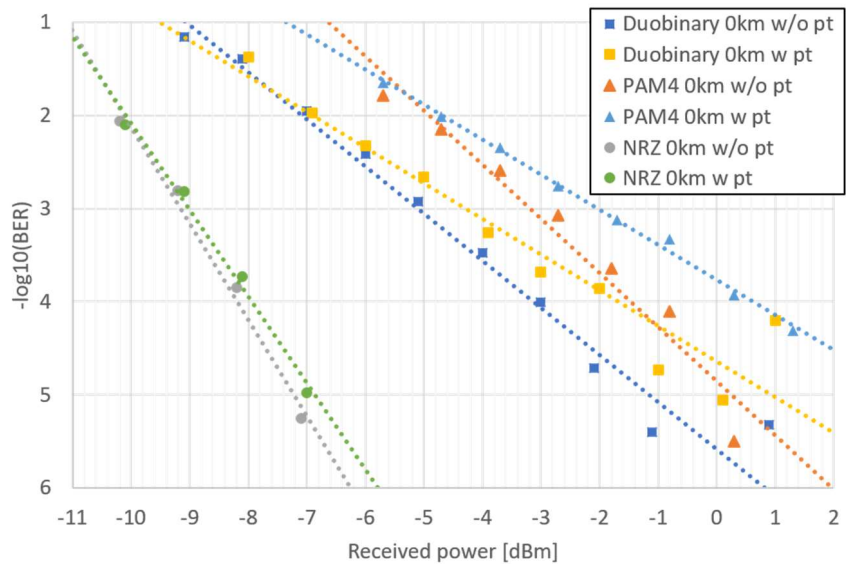

**Fig. 2:** Measured BER as function of received optical power for optical B2B.

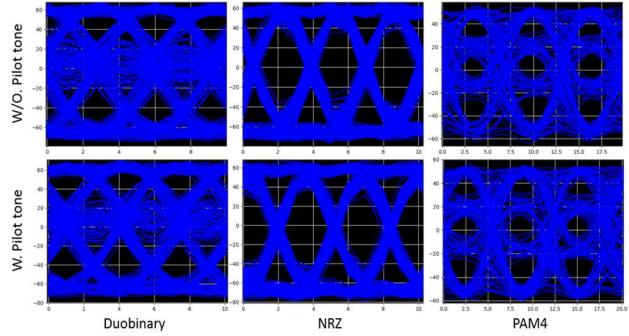

**Fig. 3:** Eyediagrams for Duobinary, NRZ and PAM4 with and without PT at highest received power.

### Results

The results from the investigation is seen in Fig. 2 in the form of BER curves. The curve shows the B2B measurements of NRZ, PAM4 and Duobinary, with and without the PT of ~8%.

It is evident from the B2B measurement, that the PT does add a receiver power penalty. This penalty increases with the number of amplitudes in the signal. The penalty is decreasing with lower received power due to overall signal attenuation. The highest penalty we observed is from PAM4 at a BER of 4.5 at the -log10 scale. Here the penalty reaches 3 dB. The results in Fig. 2 also shows that the sensitivity of the PD used is low, therefor the high receiver power. Fig. 3 shows the eyediagrams of the received signal at the highest received power. As observed from Fig. 3 the PT of 8% is not clearly sees from the eyediagrams, therefor measuring the PT modulation depth following the standard[1] is necessary.

To see the effects of the PT after transmission, PAM4 was transmitted over 20 km of SSMF, demonstrating the worst-case scenario. Fig. 4. shows the measured BER from the PAM4 transmission. As observed from Fig.4. the trend from the B2B measurements is also seen when transmission distance is increased. The penalty from the PT is observed be almost linear with received power.

### Conclusions

An experimental investigation of PT effects for multilevel partial response signals has shown the PT modulation depth of 8% can add a significant penalty (3 dB) to the received power. And further investigation should be done, so this penalty can be considered when calculating the link budget for future WDM-PONs.

### Acknowledgements

This project has received funding from the European Union's Horizon 2020 research and innovation programme under grant agreement No 762055 (project BlueSpace).
This research has been supported by the Villums Fonden through the SEES project.

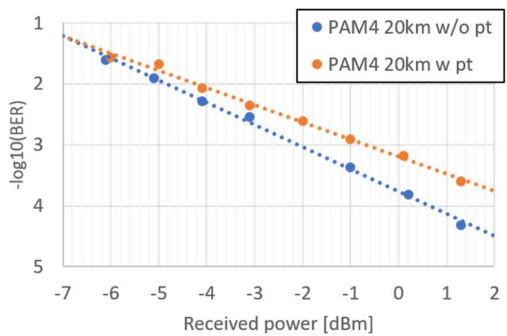

**Fig. 4:** Measured PAM4 BER as function of received power at a transmission distance of 20 km.